\documentclass[aps,prl,reprint,superscriptaddress,showpacs]{revtex4-1}

\usepackage{graphicx}
\newcommand{\bvec}[1]{\mbox{\boldmath $#1$}}

\begin{document}


\title{An Unconventional Phase Transition in BaAl$_2$O$_4$ Driven by Two Competing Soft Modes}


\author{Y. Ishii}
\email{ishii@mtr.osakafu-u.ac.jp}
\affiliation{Department of Materials Science, Osaka Prefecture University, Sakai, Osaka 599-8531, Japan.}

\author{S. Mori}
\affiliation{Department of Materials Science, Osaka Prefecture University, Sakai, Osaka 599-8531, Japan.}

\author{Y. Nakahira}
\affiliation{Department of Physics, Hiroshima University, Higashi-Hiroshima, Hiroshima 739-8526, Japan.}

\author{C. Moriyoshi}
\affiliation{Department of Physics, Hiroshima University, Higashi-Hiroshima, Hiroshima 739-8526, Japan.}

\author{H. Taniguchi}
\affiliation{Department of Physics, Nagoya University, Nagoya 464-8602, Japan.}

\author{H. Moriwake}
\affiliation{Japan Fine Ceramics Center, Nagoya 456-8587, Japan}

\author{Y. Kuroiwa}
\affiliation{Department of Physics, Hiroshima University, Higashi-Hiroshima, Hiroshima 739-8526, Japan.}


\date{\today}

\begin{abstract}
We investigated the temperature dependence of the superlattice intensity and thermal diffuse scattering intensity of BaAl$_2$O$_4$, which has a network structure with corner-sharing AlO$_4$ tetrahedra, via synchrotron X-ray diffraction experiments. 
The temperature variation of the superlattice intensity revealed that the structural phase transition occurs at $T_{\rm C} = 451.4$ K from the $P6_322$ parent crystal structure to the low-temperature superstructure with a cell volume of $2a \times 2b \times c$. 
BaAl$_2$O$_4$ exhibits an unconventional structural phase transition driven by two competing soft modes, $\bvec{q}_{\rm 1/2}\approx(1/2, 1/2, 0)$ and $\bvec{q}_{\rm 1/3}\approx(1/3, 1/3, 0)$.
When approaching the $T_{\rm C}$ from above, the soft mode with $\bvec{q}_{\rm 1/3}$ appeared first and was followed by the $\bvec{q}_{\rm 1/2}$ soft mode.
The thermal diffuse scattering intensities from both soft modes increased sharply at $T_{\rm C}$; therefore, both modes condensed simultaneously. 
The first principles calculation revealed that structural instabilities exist at the M- and K-points, at which the calculated imaginary frequencies are similar. 
The small energy difference of these structural instabilities generates the two competing soft modes and determines the eventual low-temperature crystal structure. 
\end{abstract}


\maketitle

Phonon softening, such as that in charge density waves (CDWs)\cite{TiSe2,CuxTiSe2,TaS2,LuFe2O4}, is one of the recent central issues in solid-state physics, although it has been extensively studied for decades.
Conventionally, a structural phase transition associated with soft phonons is driven by a particular soft phonon, in which the frequency of a single soft mode falls on cooling and eventually reaches zero at a transition temperature. 
However, a structural phase transition driven by more than two soft modes has never been reported thus far.

Neutron scattering, particularly the inelastic scattering technique, is a powerful tool used to study phonon dispersion and dynamics. Because the scattering intensity from phonons with wave vector $\bvec{q}$ is proportional to $1/\omega_q^2$, where $\omega_q$ is the phonon frequency, a stronger intensity is obtained from a lower phonon energy, giving rise to diffuse scatterings in reciprocal space. The observation of soft phonons is also performed within the electron diffraction or synchrotron X-ray diffraction studies \cite{Holtz}, as has been recognized thus far.

In compounds with a corner-sharing tetrahedral network structure, \it{e.g.} \rm tridymite and quartz which are SiO$_2$ modifications, characteristic diffuse scatterings have been observed in electron diffraction patterns\cite{tridymite,tridymite2,quartz2}, which are ascribed to particular phonon modes called Rigid Unit Modes (RUMs)\cite{tridymite,quartz}.
These modes are associated with the rotation or tilting of each tetrahedral block without distortion of the bonding between the corner ligand and center atom\cite{RUMs,RUMs2}. 
The distortion energy caused by these phonon modes is so small that the structure is likely to fluctuate. 
The RUM is known to occasionally act as a soft mode.
For example, nepheline, one of the derivatives of the tridymite-type structure, exhibits a structural phase transition accompanying a soft mode, which is characterized as an RUM\cite{nepheline}.

We highlighted a framework compound, BaAl$_2$O$_4$, which exhibits structural fluctuations over a wide temperature range. This compound crystallizes in a stuffed tridymite-type structure that comprises a corner-sharing, AlO$_4$ tetrahedral network with occupied six-member cavities.
The high-temperature phase with a space-group symmetry of $P$6$_3$22 undergoes a structural phase transition at approximately 400 K, accompanying the significant tilting of the AlO$_4$ tetrahedra\cite{Huang1,Huang2,Larsson}.
This results in the doubling of the cell parameters in the $a$- and $b$-axes, and the low-temperature phase with $P$6$_3$ space group exhibits a small spontaneous polarization. 
This transition has been characterized as an improper-type ferroelectric phase transition\cite{Stokes}.

RUMs have also been investigated in BaAl$_2$O$_4$ through $ab$ $initio$ calculations. According to previous studies by J.M. Perez-Mato $et$ $al$., an unstable RUM in an AlO$_4$ tetrahedral framework play an important role in the dominant structural instability of this compound\cite{Perez-Mato}.
In the electron diffraction patterns of the high-temperature phase of BaAl$_2$O$_4$, characteristic honeycomb-like diffuse streaks along three equivalent $\langle$110$\rangle$ reciprocal directions (honeycomb pattern) have been observed \cite{Abakumov}.
Because the diffuse scattering intensity is strongly dependent on the temperature, it is anticipated that the characteristic honeycomb pattern may stem from a soft mode.
Similar honeycomb patterns have also been observed for Ba$_{0.6}$Sr$_{0.4}$Al$_2$O$_4$ polycrystalline samples \cite{Fukuda}. According to the structure refinements and MEM analyses, three-site disorder exists at the bridging oxygen atoms, thus indicating that the structural phase transition of BaAl$_2$O$_4$ is of an order-disorder type.

In the present study, we performed synchrotron X-ray diffraction experiments using a BaAl$_2$O$_4$ single crystal. 
We demonstrate for the first time that the structural phase transition of the framework compound BaAl$_2$O$_4$ is unusually driven by two distinct soft modes.

Single crystals of BaAl$_2$O$_4$ were grown by the self-flux method. Previously prepared BaAl$_2$O$_4$ and BaCO$_3$ powders were mixed at a molar ratio of 50:17. The preparation method of BaAl$_2$O$_4$ powder can be found in ref.\cite{TanakaJJAP}. The mixture was placed in a platinum crucible. After heating at 1470$^{\circ}$C for 6 h, the crucible was slowly cooled to 1200$^{\circ}$C at a rate of 2$^{\circ}$C/h; then, cooled in a furnace to room temperature. The shiny, colorless crystals had a hexagonal shape edges of approximately 100 $\mu$m long and were mechanically separated from the flux. Synchrotron X-ray diffraction experiments at 300-800 K were performed at the BL02B1 beamline of SPring-8\cite{IPcamera}. The incident X-ray radiation was set at 25 keV. The diffraction intensities were recorded on a large cylindrical image-plate camera. The temperature control was performed using N$_2$ gas flow.

\begin{figure}[t]
\includegraphics[width=85mm]{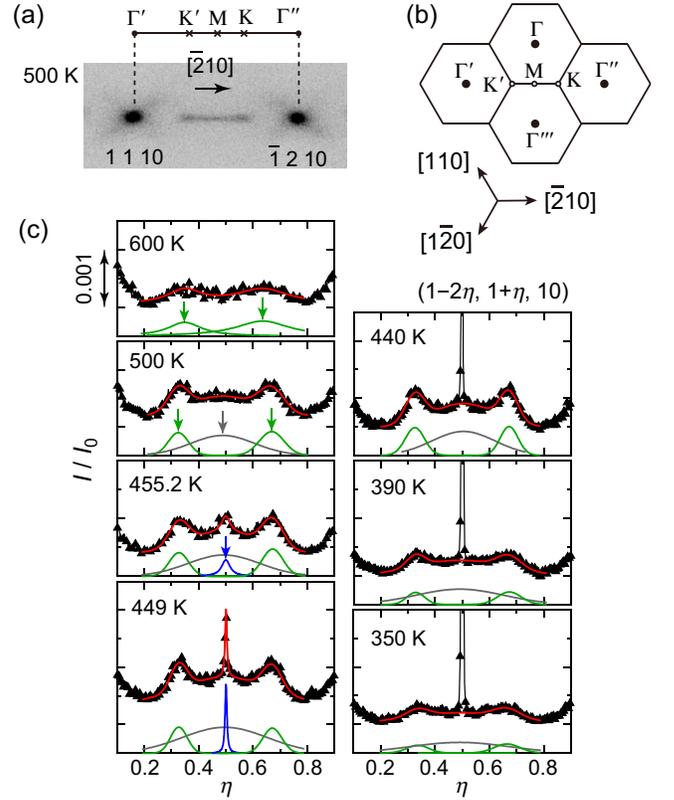}
\caption{\label{Profile} 
(Color online) (a) Typical diffraction pattern near the 11\underline{10} and $\bar{1}$2\underline{10} fundamental peaks recorded on the image-plate camera at 500 K. 
(b) Cross-sectional reciprocal space ($h$, $k$, 0) of a primitive hexagonal lattice. The symmetry points are also shown.
(c) Intensities from the 11\underline{10} to $\bar{1}$2\underline{10} fundamental peaks at 350$\sim$600 K. All profiles are drawn using the same scale. The intensities are normalized using the maximum intensity of the 11\underline{10} fundamental peak at each temperature. They are fitted using several Gaussian functions shown by green, gray and blue lines. The red lines are the superposition of these Gaussian functions. For the fitting below 445 K, the superlattice reflections have been subtracted. The green arrows indicate the peak positions of $\pm\bvec{q}_{1/3}$ diffuse scatterings. Gray and blue arrows indicate the peak positions of the broad diffuse scattering of $\bvec{q}_{1/2}$ and the superlattice reflection, respectively. Data were corrected upon heating.}
\end{figure}

Figure 1(a) shows the diffraction pattern near the 11\underline{10} and $\bar{1}$2\underline{10} fundamental peaks obtained at 550 K. 
Thermal diffuse scatterings are observed along [$\bar{2}$10], which is one of the three equivalent $\langle$110$\rangle$ directions. 
This direction corresponds to $\Gamma$'-K'-M-K-$\Gamma$" in the reciprocal space, as shown in Fig. 1(b).
The intensities of these diffuse scatterings were strongly dependent on the temperature, as reported in ref. \cite{Abakumov}, and existed at all temperatures used for the measurements.
In addition, cross-shaped thermal diffuse scatterings were observed around each fundamental peak, although the intensities of these diffuse scatterings were largely independent of the temperature and were present at all temperatures used for the measurements.

We investigated the temperature dependence of the diffuse scatterings between two fundamental peaks.
Figure 1(c) shows the intensity obtained at typical temperatures at $(1-2\eta,1+\eta,10)$, in which $\eta = 0$ and $\eta = 1$ correspond to the (1,1,10) and ($\bar{1}$,2,10) reciprocal points, respectively. 
All of the profiles obtained at various temperatures between 300 and 800 K are shown in Figs. S2 and S3.
The intensities were normalized using the 11\underline{10} fundamental peak at each temperature.
The intensities below $\eta=0.2$ and above $\eta=0.8$ come from the 11\underline{10} and $\bar{1}$2\underline{10} fundamental peaks.
These diffuse scatterings within $0.2<\eta<0.8$ can be satisfactorily fitted using several Gaussian functions.
The red line shows the superposition of these fitting curves.
For fitting below 445 K, the strong superlattice reflections were subtracted from each profile.

At 600 K, there are two broad diffuse peaks near $\eta=1/3$ and $2/3$, as indicated by the green curves.
Similar peaks are also observed at 700 and 800 K.
In addition to these two diffuse peaks, another broad diffuse peak can be seen near $\eta=1/2$ at 500 K, as indicated by the gray curve. 
At 455.2 K, an additional small peak appears at $\eta=1/2$, as indicated by the blue curve. This small peak developed as a superlattice reflection at low temperature. 
As described later, the diffuse peaks near $\eta=1/2$ and $1/3$ can be attributed to the two soft modes with the wave vectors of $\bvec{q}\approx(1/2, 1/2, 0)$ and (1/3, 1/3, 0).
Hereafter, we call these scatterings the $\bvec{q}_{1/2}$ and $\bvec{q}_{1/3}$ diffuse scatterings, respectively. 
The diffuse scattering near $\eta=2/3$, which is equivalent to that near $\eta=1/3$, is denoted as the $-\bvec{q}_{1/3}$ diffuse scattering.
These diffuse scatterings of $\bvec{q}_{1/2}$ and $\pm\bvec{q}_{1/3}$ existed at all temperatures used for the measurements.
The small peak at $\eta=1/2$ at 455.2 K, indicated by the blue arrow in Fig. 1 (c), can be understood as the short-range ordering of the $2a \times 2b \times c$ superstructure.

\begin{figure}[t]
\includegraphics[width=85mm]{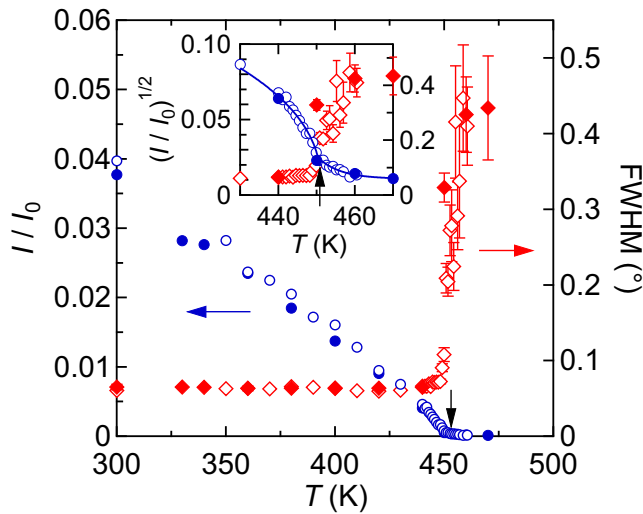}
\caption{\label{SuperIntFWHM} (Color online) Normalized maximum superlattice intensity (left axis) and FWHM (right axis) plotted against temperature. The open and closed symbols represent the data obtained upon heating and cooling, respectively. The standard deviations of the normalized intensity data were considerably smaller than the marker size. The inset shows the square root of the superlattice intensity (left axis) and FWHM (right axis) near $T_{\rm C}$. $T_{\rm C}$ values are indicated by arrows in each figure. The solid lines are the fits using a power law ($T<T_{\rm C}$) and an exponential function ($T>T_{\rm C}$).}
\end{figure}

Figure 2 shows the maximum intensity and full width at half maximum (FWHM) of the superlattice reflections as a function of temperature. 
The small peaks such that observed at 455.2 K in Fig. 1 (c) are also included.
The inset shows the square root of the superlattice intensity, $(I/I_0)^{1/2}$, and the FWHM in the range of $T=$ 430-470 K.
The superlattice intensity decreases with increasing temperature.
It shows a kink at approximately 450 K, as indicated by the arrows in the figures. Above $\sim$450 K, the weak intensity still exists as a tail, and the FWHM increases abruptly with increasing temperature. 
A similar tail behavior has also been observed in BaMnF$_4$, with high concentrations of defects\cite{Ryan,Cox}, and manganese compounds\cite{ShimomuraPRL,ShimomuraPRB,ShimomuraJPSJ}.
Below $\sim$450 K, the long-range ordering of the $2a \times 2b \times c$ superstructure appears. Therefore, it is reasonable to regard this temperature as $T_{\rm C}$.
The $(I/I_0)^{1/2}$ values shown in the inset were fitted using a power function and exponential function.
The curve below $T_{\rm C}$ was fitted using a power function, $A_0(T_{\rm C}-T)^{\beta}$, where $\beta=0.40$ and $T_{\rm C}$ = 451.4. $A_0$ is a coefficient. The obtained critical index ${\beta}$ coincides well with the theoretical value of the order-disorder type\cite{Gebhardt}. An exponential function was found to accurately reproduce the tail part above $T_{\rm C}$.

\begin{figure}[t]
\includegraphics[width=85mm]{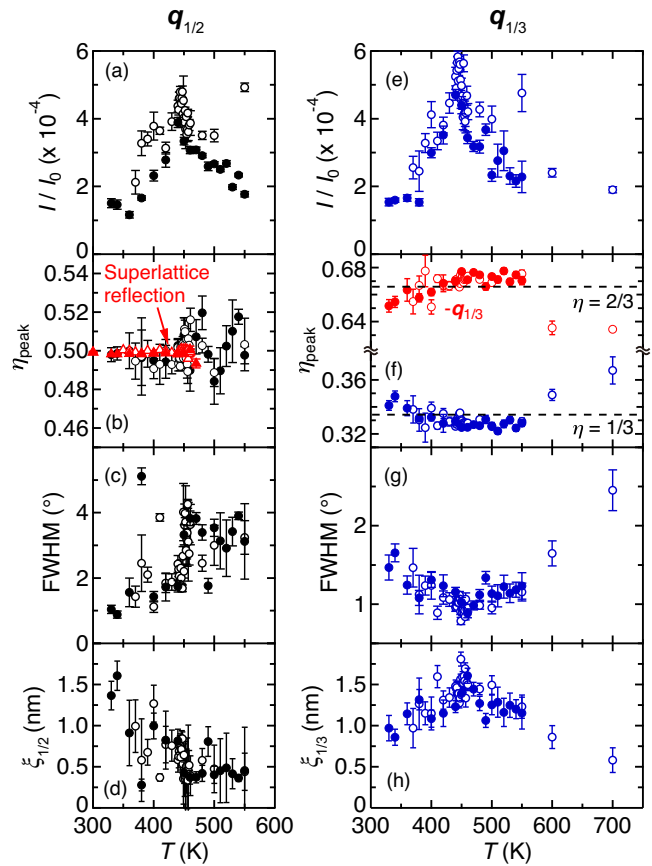}
\caption{\label{Fig3-Diffuse} (Color online) Normalized maximum intensity, peak position ($\eta_{\rm peak}$), FWHM, and correlation length ($\xi$) of the thermal diffuse scatterings of the $\bvec{q}_{1/2}$ and $\bvec{q}_{1/3}$ soft modes. (a)-(d) and (e)-(h) correspond to the thermal diffuse scatterings of the $\bvec{q}_{1/2}$ and $\bvec{q}_{1/3}$ soft modes, respectively. The peak positions of the superlattice reflections and the $-\bvec{q}_{1/3}$ thermal diffuse peaks are also plotted in (b) and (f), respectively. The open and closed symbols represent the data obtained upon heating and cooling, respectively.}
\end{figure}

Figures 3(a)$\sim$(h) display the temperature dependence of the intensities, peak positions ($\eta_{\rm peak}$), FWHMs the correlation length ($\xi$) of the $\bvec{q}_{1/2}$ and $\pm\bvec{q}_{1/3}$ thermal diffuse scatterings. 
The scattering intensities of both the $\bvec{q}_{1/2}$ and $\bvec{q}_{1/3}$ diffuse scatterings increased sharply at $T_{\rm C}$, as shown in Figs. 3 (a) and (e). 
This result means that these diffuse scatterings stem from soft modes, and both modes condense at $T_{\rm C}$ simultaneously. 
With respect to the $\eta_{\rm peak}$, the $\bvec{q}_{1/2}$ diffuse peak is sufficiently broad that the $\eta_{\rm peak}$ of the $\bvec{q}_{1/2}$ diffuse peak is difficult to determine with high accuracy, as shown in Fig. 3 (b). 
It is reasonable to consider that the $\bvec{q}_{1/2}$ diffuse peak is at the commensurate value of $\eta=1/2$. 
The peak positions of the superlattice reflections are also plotted in this figure.
The superlattice reflections are at the commensurate position.
In contrast, the $\eta_{\rm peak}$ of the $\bvec{q}_{1/3}$ diffuse peak shows an intriguing behavior, as shown in Fig. 3 (f); namely, $\eta_{\rm peak}\sim0.36$ above 600 K, $\eta_{\rm peak}\sim0.32$ at $T_{\rm C}<T<550$ K, and then, it shifts toward 1/2 below $T_{\rm C}$.
Neither $\eta_{\rm peak}$ above $T_{\rm C}$ nor below $T_{\rm C}$ show the commensurate values.
Similar results were obtained for the $-\bvec{q}_{1/3}$ diffuse peak. 
The FWHMs of the $\bvec{q}_{1/2}$ and $\bvec{q}_{1/3}$ diffuse peaks are plotted in Figs. 3(c) and (g), respectively. The FWHM of the $\bvec{q}_{1/2}$ diffuse peak abruptly decreases at $T_{\rm C}$ as the superlattice develops. 
Correspondingly, the $\xi_{1/2}$ develops below $T_{\rm C}$.
In contrast, the FWHM of the $\bvec{q}_{1/3}$ diffuse peak decreases gradually with decreasing temperature, showing a minimum at $T_{\rm C}$, and then increases below $T_{\rm C}$.
Consequently, $\xi_{1/3}$ exhibits a maximum at $T_{\rm C}$, as shown in Fig. 3 (h). 

The results shown in Figs. 3 indicate that the two soft modes of the $\bvec{q}_{1/2}$ and $\bvec{q}_{1/3}$ wave vectors coexist and compete above $T_{\rm C}$. 
In the high-temperature phase, the $\bvec{q}_{1/3}$ soft mode has already appeared above 600 K, as shown in Fig. 3 (e), whereas the $\bvec{q}_{1/2}$ soft mode appears at 550 K. 
Moreover, the $\bvec{q}_{1/3}$ soft mode exhibits a stronger intensity and longer $\xi_{1/3}$ than the $\bvec{q}_{1/2}$ soft mode. 
Nevertheless, the $\bvec{q}_{1/2}$ soft mode is selected to form the superstructure below $T_{\rm C}$, and the $\bvec{q}_{1/3}$ soft mode is overcome by the later-developed $\bvec{q}_{1/2}$ soft mode.
To the best of our knowledge, this structural phase transition has never been reported previousely. 
In this unprecedented structural phase transition, the soft phonon fluctuates between the two wave vectors of $\bvec{q}_{1/2}$ and $\bvec{q}_{1/3}$, that is, ``fluctuation in phonons" in the $k$-space.

We investigated the phonon dispersions via first principles calculations. 
First-principles calculations were performed using the projector-augmented wave (PAW) method\cite{PAW} within the framework of density functional theory (DFT)\cite{DFT1,DFT2}, as implemented in VASP code\cite{VASP1,VASP2}. Exchange-correlation interactions were treated by the generalized gradient approximation (GGA)\cite{GGA}. Lattice constants and internal coordinates were considered fully optimized when the residual Hellmann-Feynman forces were smaller than $1.0\times10^{-3}$ eV/\AA. Phonon dispersion calculations were performed using the Phonon code\cite{PHONOPY}.
The calculated phonon dispersion of BaAl$_2$O$_4$ is shown in Fig. 4. 
We found one of the acoustic phonon branches showing the imaginary frequencies at the M-, K- and A-points, which indicates the structural instability giving rise to the structural phase transition.
The calculated imaginary frequencies were 1.80, 1.77 and 0.67 THz at the M-, K- and A-points, respectively.
The structural instabilities at the M- and K-points are larger than that at the A-point. 
These two structural instabilities cause the $\bvec{q}_{1/2}$ and $\bvec{q}_{1/3}$ soft modes.
These modes are related to the previously reported RUMs\cite{Perez-Mato}.
The calculation also reveals that the difference in the destabilization energy at the K-point and M-point is quite small. 
This subtle difference determines the eventual crystal structure of the low-temperature phase.
The previously reported three-fold superstructure in Ba(Al,Fe)$_2$O$_4$\cite{Fe-BaAl2O4} is probably caused by the slight modification in the structural stability owing to the Al-site disorder. 
\begin{figure}[t]
\includegraphics[width=70mm]{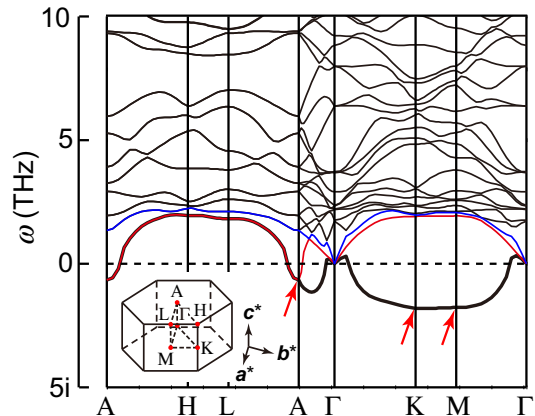}
\caption{\label{Fig4-Phonon} (Color online) Phonon dispersion curves of BaAl$_2$O$_4$ obtained by first principles calculations. The calculations were performed on the $P6_322$ parent structure. The inset shows the first Brillouin zone of a primitive hexagonal lattice. Three acoustic phonon branches are shown as red, blue, and thick black lines. One of the acoustic phonon branches shows imaginary frequencies at the A-, K-, and M-points, as indicated by red arrows.}
\end{figure}

Among the three acoustic modes, the dispersion indicated by the blue line is the longitudinal acoustic (LA) mode.
The two other modes are the transverse acoustic (TA) modes. 
The TA modes degenerate at the Brillouin zone boundary of $k_z = \pi/c$, although the degeneracy is lifted at $k_z=0$, that is, the mode that acts as the soft mode is one of the TA modes.
Because the diffuse scattering intensity is proportional to $|\bvec{K}\cdot\bvec{e}_{j,q}|/\omega_q^2$, where $\bvec{K}$ and $\bvec{e}_{j,q}$ are the scattering vector and atomic displacement unit vector of the $j$ atom, respectively, the scattering intensity by the transverse soft mode with $\bvec{q} // [110]$ is expected to disappear along $\bvec{K} // [110]$. This is evidenced by the TEM experiments shown in Fig. S4.

In conclusion, BaAl$_2$O$_4$ is a unique compound that shows a structural phase transition driven by the two distinct soft modes of $\bvec{q}_{1/2}$ and $\bvec{q}_{1/3}$. 
Its high-temperature phase has competing structural instabilities at the M- and K-points, which generate the $\bvec{q}_{1/2}$ and $\bvec{q}_{1/3}$ soft modes, respectively. 
The two modes condense together at $T_{\rm C}$, which results in the development of strong thermal diffuse scatterings along the K'-M-K line. 
The subtle energy difference between the modes determines the eventual crystal symmetry of the low-temperature phase.
This characteristic feature can be attributed to the network structure of the corner-sharing AlO$_4$ tetrahedra in this compound.

\begin{acknowledgments}
This work was partially supported by a Grant-in-Aid for Scientific Research from the Ministry of Education, Culture, Sports, Science and Technology of Japan (MEXT), MEXT Element Strategy Initiative Project, Grant-in-Aid for Challenging Exploratory Research (15K14120), and Grant-in-Aid for Scientific Research on Innovative Areas ``Nano Informatics'' (25106008) from Japan Society for the Promotion of Science (JSPS). The experiments at SPring-8 were performed with the approval of the Japan Synchrotron Radiation Research Institute (JASRI; Proposal Nos. 2014A0078, 2014A1323, 2014B0078, and 2015A1507).
\end{acknowledgments}


\end{document}